\documentclass{article}

\usepackage{booktabs}
\usepackage[table,xcdraw]{xcolor}

\usepackage[preprint, nonatbib]{neurips_2022}

\usepackage[utf8]{inputenc} 
\usepackage[T1]{fontenc}    
\usepackage{hyperref}       
\usepackage{url}            
\usepackage{booktabs}       
\usepackage{amsfonts}       
\usepackage{nicefrac}       
\usepackage{microtype}      
\usepackage{xcolor}         

\usepackage{biblatex}

\usepackage{tabularx}
\usepackage{seqsplit}

\newcolumntype{L}{>{\raggedright\arraybackslash}X}
\newcommand{\wrapitems}[1]{%
  \seqsplit{}\wrapitem{}#1%
}
\newcommand{\wrapitem}[1]{#1}

\title{Scientists’ Perspectives on the Potential for Generative AI in their Fields}

\author{%
  Meredith Ringel Morris \\
  Google Research\\
  \texttt{merrie@google.com} \\
}

\begin{document}

\maketitle

\begin{abstract}
  Generative AI models, including large language models and multimodal models that include text and other media, are on the cusp of transforming many aspects of modern life, including entertainment, education, civic life, the arts, and a range of professions. There is potential for Generative AI to have a substantive impact on the methods and pace of discovery for a range of scientific disciplines. We interviewed twenty scientists from a range of fields (including the physical, life, and social sciences) to gain insight into whether or how Generative AI technologies might add value to the practice of their respective disciplines, including not only ways in which AI might accelerate scientific discovery (i.e., research), but also other aspects of their profession, including the education of future scholars and the communication of scientific findings. In addition to identifying opportunities for Generative AI to augment scientists’ current practices, we also asked participants to reflect on concerns about AI. These findings can help guide the responsible development of models and interfaces for scientific education, inquiry, and communication. 
  
\end{abstract}

\section{Introduction}

The past few years have seen enormous leaps forward in the capabilities of AI (Artificial Intelligence) technologies, particularly in the successive introductions of Generative AI systems that ingest huge quantities of data and learn statistical patterns that enable the models to generate novel content, typically based on a natural-language input (\textit{prompts}). These models can produce a range of content types including text \cite{gpt3_paper, gpt4techreport, lamda_arxiv, palm}, images \cite{dalle2, imagen, partiscaling}, video \cite{phenaki, imagenvideo}, and audio \cite{musiclm, audiolm, speechSynth}. The increasing and novel capabilities of these AI systems is poised to transform many aspects of modern life; in this paper, we specifically consider how Generative AI advances might transform the practice of science, including the life sciences, physical sciences, and social sciences. 

We interviewed twenty scientists from a range of fields to gain insight into whether or how Generative AI technologies might add value to the practice of their respective disciplines, including not only ways in which Generative AI might accelerate scientific discovery (i.e., research), but also other aspects of their profession, including the education of future scholars and the communication of scientific findings to other scientists and to the general public. Through a qualitative analysis of the interviews, we identified a common set of themes among scientists’ understanding of how new AI technologies might benefit their profession, as well as concerns around risks such technologies may introduce. 

Our participants foresaw the potential for both positive and negative impacts of Generative AI. Potential benefits include novel educational paradigms, AI-supported literature review including cross-disciplinary connections and insights, tools to accelerate dataset creation and cleaning, AI-assisted coding to accelerate various aspects of experimentation and analysis, novel AI-assisted methodologies for data collection and experimental design, and AI-enhanced writing and peer-review systems. Concerns included AI-aided cheating and lost opportunities for critical reflection, the introduction of bias through the use of AI, factuality and trustworthiness of AI tools, the proliferation of publication spam, fake data, and scientific misinformation. This work contributes findings that can support more human-centered and responsible development of AI tools for the sciences, by representing the perspectives of experts with deep knowledge of the practice of various disciplines within the physical, life, and social sciences.   

In this article, we first discuss related work on Generative AI and on AI-assisted science. We then describe our participants’ backgrounds and our interview method. Then, we summarize the key themes that emerged from our research, covering the impacts of Generative AI on scientific education, dataset creation, literature reviewing, coding, discovery, and communication. We conclude by discussing how scientists consider the issue of trust with respect to any AI tool they might use in their work, and reflecting on our findings, their limitations, and avenues for future exploration.

\section{Related Work}

This investigation builds on human-centered approaches to understanding the potential for and impacts of AI on particular end-user communities through engagement with expert stakeholders, such as software developers \cite{dibia2022aligning, bird2023programming, mozannar2022reading, pairSoftware}, medical professionals \cite{medicalOnboarding, medicalImaging}, artists \cite{caiAImusic, chang2023prompt, kulkarni2023word}, etc. As this work focuses particularly on the applicability of emerging Generative AI technology to scientific fields, our literature review focuses on the state of the art in Generative AI and on AI-assisted scientific inquiry. 

\subsection{Generative AI}
The introduction of the transformer neural network architecture in 2017 \cite{NIPS2017_3f5ee243} introduced a new era in machine learning, enabling the creation of a variety of powerful pre-trained, generative AI models (sometimes called \textit{foundation models} \cite{foundation_models}) capable of producing novel content in several modalities. While generative AI initially took off in the language domain (i.e., the GPT family of language models \cite{gpt3_paper, gpt4techreport} and their competitors \cite{lamda_arxiv, palm}), models now also exist for a variety of other modalities including imagery \cite{dalle2, imagen, partiscaling}, video \cite{phenaki, imagenvideo}, and audio \cite{musiclm, audiolm, speechSynth}. \footnote{Note that due to the recent explosion in advances in generative AI, we do not attempt to provide an exhaustive list of models or their capabilities, but rather to cite some noteworthy examples from each class of technology to illustrate the state-of-the-art.} These models provide novel opportunities for human-computer interaction and human-AI collaboration \cite{merriedesignspace}, including by augmenting the capabilities of scientists. 

While impressive in their (quickly increasing) capabilities (there is debate, for instance, as to whether GPT-4 may be a step toward Artificial General Intelligence \cite{sparksofagi}), Generative AI still has many limitations, including potential risks such as hallucinations (fabrications, non-factual outputs, e.g. \cite{dziri2022origin}), production of biased or offensive content, memorization of training data, and ethical concerns over the provenance of training materials \cite{parrots, deepmind_llm_ethics}. In our study, we are mindful to consider not only the opportunities that Generative AI will create for enhancing scientific education, discovery, and communication, but also the challenges and risks it may simultaneously entail. 

\subsection{AI-Assisted Science}

While the application of the emerging generation of Generative AI tools to the sciences is nascent (and the subject of the study presented in this paper), prior generations of AI such as ML-based modeling, optimization, data analysis, and earlier deep learning techniques have accelerated discovery across a range of scientific disciplines as well as spawning new sub-disciplines that prioritize computational methodologies such as modeling and simulation over traditional techniques such as empirical work in the lab or field.\footnote{Note that the volume of AI-assisted scientific endeavors are too large to list exhaustively; we focus on a few noteworthy and representative examples to give the reader a sense of the space.} 

One of the most prominent recent examples of the impact of AI on scientific discovery is DeepMind's AlphaFold project \cite{alphafold}, which uses ML to predict the three-dimensional structure of proteins from an amino acid sequence, thereby accelerating biology research that formerly required time-consuming in-lab protein synthesis. In only the past decade, AI has supported novel discoveries and methods across a range of scientific disciplines, including mathematics \cite{deepmindMath},  astronomy \cite{deepmindAstronomy, Jin_2022}, climate science \cite{googleRainPrediction}, conservation biology \cite{norouzzadeh2019deep}, biochemistry \cite{Goldman2022.12.30.522318}, medical diagnostics \cite{rajpurkar2017chexnet}, drug discovery \cite{Ekins2019, Melo2021}, and myriad others.

Some scientists have already begun to explore the potential of Generative AI within their fields, either by utilizing off-the-shelf models or fine-tuning or training models specifically for their domains, such as the ClimaX model for weather prediction \cite{nguyen2023climax} or GenSLM \cite{geneLLM} (a language model adapted for genomic research). In 2022 Google  introduced a language model, Minerva, that exhibited success on benchmarks of undergraduate-level quantitative reasoning problems in a range of scientific domains \cite{minerva}. In November 2022, Meta released Galactica \cite{taylor2022galactica}, a generative language model trained on a scientific corpus, though they pulled the system from public use after only a few days after it generated false scientific citations and offensive pseudo-scientific research abstracts. 

Interest in the potential for Generative AI to accelerate science is growing, e.g. with propositions of grand challenges such as the Nobel Turing Challenge to "develop a highly autonomous AI system that can perform top-level science, indistinguishable from the quality of that performed by the best human scientists, where some of the discoveries may be worthy of Nobel Prize level recognition and beyond" \cite{Kitano2021}. The National Academy of Sciences has also recently considered the implications of modern AI for scientific discovery \cite{NAP26532}, as has the 2023 Stanford AI Index report \cite{stanfordAI2023}, which devotes an entire chapter to reflecting on recent advances in AI for Science, including in domains such as nuclear fusion \cite{Degrave2022}, matrix manipulation \cite{Fawzi2022}, chip design \cite{prefixRL}, and antibody design \cite{Shanehsazzadeh2023.01.08.523187}. 

Our work builds on these efforts by taking a human-centered approach that directly seeks and synthesizes perspectives from experts from the physical, life, and social sciences into the potential benefits and pitfalls of generative AI for education, discovery, and communication in their fields of speciality; these findings can inform the design of novel AI tools and human-AI collaboration interfaces and paradigms aligned to scientists' (and society's) goals. 

\section{Method}

\subsection{Participants}
We interviewed twenty scientists (seven female, thirteen male), all of whom had Ph.D. degrees in a field relevant to their discipline. We included participants with backgrounds in the physical sciences (e.g., physics, geology, chemistry), the life sciences (e.g., immunology, bioinformatics, ecology), and the social sciences (e.g., sociology, anthropology, behavioral economics). Our interview specifically excluded participants with a doctorate in computer science, since knowledge of the practice of computer science is widespread among those developing Generative AI technologies, and the goal of this work was to obtain broader perspectives. 

For confidentiality reasons, all participants were affiliated with Alphabet; nine were full-time Alphabet employees, and eleven were tenure-track or research faculty at major universities or research institutes with part-time roles at Alphabet (e.g., consulting one-day-per-week, full-time for a fixed-term sabbatical, etc.). 

We recruited participants using snowball-sampling techniques to identify scientists affiliated with Alphabet and by asking Alphabet’s University Relations department to provide a list of visiting faculty (e.g., part-time consultants, sabbatical visitors, etc.) with Ph.D. degrees in fields besides computer science. 

Table \ref{participant-table} summarizes the disciplinary and demographic background of the twenty interviewees.

\begin{table}[ht]
\centering
\begin{tabularx}{\textwidth}{@{}lLLl@{}}
\toprule
\textbf{ID}  & \textbf{Discipline} & \textbf{Affiliation} & \textbf{Gender} \\ \midrule
P1  & \wrapitems{Behavioral and Experimental Economics; Market Theory} & \wrapitems{Professor of Economics (primary); Alphabet (part-time)} & M \\
\hline
P2 & \wrapitems{Earthquake Science; Geophysics} & \wrapitems{Prof. of Earth \& Planetary Sciences (primary); Alphabet (part-time)} & M \\
    \hline
    P3 & \wrapitems{Immunology; Genomics} & \wrapitems{Alphabet (primary); Adjunct Prof. in Dept. of Medicine (part-time)} & M \\
    \hline
    P4 & \wrapitems{Materials Science} & \wrapitems{Alphabet (primary)} & F \\
    \hline
    P5 & \wrapitems{Physics} & \wrapitems{Alphabet (primary)} & M \\
    \hline
    P6 & \wrapitems{Hydrology} & \wrapitems{Alphabet (primary)} & M \\
    \hline
    P7 & \wrapitems{Biochemistry} & \wrapitems{Prof. of Biochemistry (primary); Alphabet (part-time)} & F \\
    \hline
P8 & \wrapitems{Electrical Engineering; Physics} & \wrapitems{Prof. of Aerospace Engineering (primary); Alphabet (part-time)} & F \\
    \hline
    P9 & \wrapitems{Climate Science} & \wrapitems{Alphabet (sabbatical); Prof. of Geophysics (primary)} & M \\
    \hline
    P10 & \wrapitems{Quantum Chemistry} & \wrapitems{Prof. of Chemistry (primary); Alphabet (part-time)} & M \\
    \hline
P11 & \wrapitems{Communications Science; Speech-Language Pathology} & \wrapitems{Prof. of Communications Sciences \& Disorders (primary); Alphabet (part-time)} & M \\
    \hline
    P12 & \wrapitems{Quantum Computing Hardware} & \wrapitems{Alphabet (primary)} & M \\
    \hline
    P13 & \wrapitems{Sociology} & \wrapitems{Alphabet (primary)} & M \\
    \hline
    P14 & \wrapitems{Sociology; Science \& Technology Studies} & \wrapitems{Alphabet (primary)} & F \\
    \hline
    P15 & \wrapitems{Seismology} & \wrapitems{Prof. of Earth \& Planetary Science (primary); Alphabet (part-time)} & M \\
    \hline
P16 & \wrapitems{Physics; Quantum Algorithms} & \wrapitems{Alphabet (primary)} & M \\
    \hline
    P17 & \wrapitems{Medical Anthropology} & \wrapitems{Alphabet (primary)} & F \\
    \hline
    P18 & \wrapitems{Stem Cell Biology; Bioinformatics} & \wrapitems{Scientist at a nonprofit Biology research institute (primary); Alphabet (part-time)} & F \\
    \hline
    P19 & \wrapitems{Physics; Mathematics; Fluid Dynamics} & \wrapitems{Prof. of Engineering \& Geological Sciences (primary); Alphabet (part-time)} & M \\
    \hline
    P20 & \wrapitems{Geography} & \wrapitems{Research Prof. of Forestry (primary); Alphabet (part-time)} & F \\

\bottomrule 
\end{tabularx} 
\caption{\label{participant-table}All interview participants were either full-time or part-time employees of Alphabet; part-time employees had primary affiliations at universities or research institutes. University names have been removed for anonymity, but include major U.S. research institutions such as Harvard University, Stanford University, U.C. Berkeley, and CalTech. All participants were employed in the United States at the time of the interview. All participants had a Ph.D. (doctorate) degree in a discipline related to their field of study.}
\end{table}

\subsection{Interview Structure}
A single researcher conducted all twenty interviews over a monthlong period spanning late January and early February 2023, using a semi-structured interview technique, which used a basic script for questions and allowed time for following up with spontaneous, in-depth questions on topics that emerged during the course of the interview. Participants were reminded that they were free to skip any questions or ask for clarification if needed. Each interview lasted between thirty and forty-five minutes. 

Interviews were conducted over video call, using Google Meet. The researcher took interview notes in real-time and also recorded the sessions in case there were gaps in the notes (four sessions were not recorded due to researcher error). Participants were not offered a gratuity; they volunteered to participate based on their interest in helping Alphabet responsibly innovate in AI for scientific professions. 

The interview had four primary categories of questions. First, we collected background information (e.g., demographics, role). Second, we asked questions to learn more about the participants’ scientific discipline, to understand more about specific areas of inquiry and methodologies. Third, we introduced the topic of Generative AI; after probing participants’ familiarity with state-of-the-art, publicly available models (e.g., ChatGPT, Dall-E 2), we asked them a series of questions envisioning whether or how generative AI might apply to various career stages (educating future scientists, conducting research, disseminating findings) and various aspects of the scientific method (choosing research problems, forming hypotheses, designing experiments, collecting data, analyzing data). Finally, we asked a set of questions probing potential concerns or downsides of the use of Generative AI in their field.

The interview script was as follows:

\textbf{Background Questions}
\begin{itemize}
    \item Consent to participate in the interview and to record the video call
    \item Demographic information
    \item Field of study
    \item Affiliation with Alphabet and any additional affiliations
\end{itemize}

\textbf{Understanding the Discipline}
\begin{itemize}
    \item Give an example of a scientific challenge you are working on, in three sentences or less so that a lay person can understand. 
    \item What do you think are the biggest challenges in your field
    \item What do scientists/faculty in your field spend time on that you wish could be automated?
    \item What is a challenge that PhD students/postdocs/junior scholars in your field face?
\end{itemize}

\textbf{Brainstorming about Generative AI}
\begin{itemize}
    \item Are you familiar with generative AI models (language models such as GPT-3 or ChatGPT, image models such as Dall-E 2, etc.)?  If so, explain your level of experience with these models. Have you ever used such a model? \textit{(If not, interviewer explains what generative AI is.)}
    \item Can you envision ways in which Generative AI tools (either current tools or future, more powerful tools) might impact: 
    \begin{itemize}
        \item The education of undergraduate or graduate students in your field?
        \item Learning about related work in your field or from other disciplines?
        \item The way in which you identify research questions or develop or test hypotheses?
        \item Practices for experimental design, data collection, or analysis?
        \item The communication or dissemination of research findings?
    \end{itemize}
\end{itemize}

\textbf{Concerns about AI}
\begin{itemize}
    \item What would be required for you to trust a Generative AI tool used in the sciences?
    \item What concerns do you have about potential uses of Generative AI in your field or in science in general?
    \item Any additional comments you’d like to share?
\end{itemize}

\subsection{Qualitative Data Analysis}
Interview notes (and the original recordings which were automatically transcribed by AI software, as necessary) were analyzed using qualitative data analysis techniques. All analyses were performed by a single researcher (the same researcher who conducted the interviews), with more than twenty years of professional experience in qualitative methods. 

The researcher employed an inductive analysis \cite{inductivequal} with open coding \cite{groundedtheory}, iteratively reviewing the interview notes and transcripts to develop a set of themes. The researcher then used these themes to conduct an affinity diagramming analysis \cite{affinitydiagram}, mapping responses to themes (and merging/refining some thematic categories as a result). The Findings section is organized around these emergent themes, and uses responses from participants to illustrate each theme. Quotation marks indicate verbatim quotes, whereas other responses are summarized. 

\section{Findings}
Several themes permeated our participants’ responses. Interviewees noted a range of both opportunities and concerns regarding Generative AI across several aspects of scientific work. In the following sections, we discuss each of the major themes that arose in our interviews, which include education, data, literature reviews, coding, discovery, and communication. We also discuss the topic of how the use of Generative AI may impact trust in the quality of scientific research. 

\subsection{Education}
Our interviewees agreed that advances in Generative AI were likely to change many aspects of both undergraduate- and graduate-level education in the sciences. Many of these changes are potentially positive, such as innovation in instructional methods and lowering barriers for English language learners. On the other hand, many participants also expressed concerns about the impacts of AI-assisted cheating on evaluation methods and critical thinking. 

\subsubsection{Education: Opportunities for Generative AI}
Participants saw many opportunities for Generative AI to enhance science education. Generative models could be used to create new, engaging lesson plans. For example, a professor could ask a language model to suggest ideas for real-life examples to use in lecture to illustrate a theorem or for a list of homework questions that emphasize a particular concept or for other sets of inspirational ideas that the faculty member could choose from in order to develop or update lectures, lab activities, homework, or exams. AI models might also support developing new types of demos or interactive experiences for students; for instance, P2 described how he already uses OpenAI’s generative image model DALL-E 2 for an undergraduate course on earth sciences by asking the model to create speculative illustrations of how parts of the world might look after natural disasters (e.g., “Boston after an earthquake”); such artifacts can enrich discussion and debate during class. 

Others saw opportunities for language models to yield a new class of intelligent tutoring systems. Such tools might provide students feedback on their homework, giving detailed explanations of why a student’s initial answer was wrong and helping them arrive at a correct solution on their own (rather than “cheating” by automatically rewriting the homework for them). P11 suggested that language models could be used by students to engage in reflective Q\&A in order to test their own comprehension of scientific concepts - students could ask a chatbot questions to learn about a concept, and then explain the concept back to the bot to test the depth of their understanding and grow their mental model. P16 similarly saw value in AI bots that would have pedagogical conversations with students to help them self-test their knowledge of a topic.

Language models could also help extend the limited time resources of both faculty and teaching assistants (TAs). Students could ask a chatbot questions as a first option when questions arose, and only need to visit TA or faculty office hours if the chatbot were unable to help sufficiently. P10 observed that not only could students ask content-oriented questions to a chatbot before falling back on TA hours, but that chatbots could also answer process-oriented questions, such as how to troubleshoot mathematical, statistical, or simulation software often needed to tackle college-level science problem sets. P4 noted that being able to ask questions about advanced concepts to a scientific chatbot (e.g., “Why are topological insulators a hot field?”) could provide access to information for students whose questions might otherwise go unanswered (if they do not have easy access to an expert in a specialized field). 

Scientific jargon can be off-putting to students, and can be an obstacle to learning. P16 identified an opportunity for language models to reduce this friction by allowing students to ask a language model to explain to them the meanings of particular terms or concepts in order to improve comprehension of course materials. P7 also noted that her field (biochemistry) has a huge “jargon barrier” and that there is an opportunity for language models not only to support comprehension of jargon in reading materials, but also in the classroom itself, such as when a seminar presenter mentions a particular protein and the student needs to get a quick reminder of that protein’s properties and interactions in order to appreciate the rest of the lecture. 

Finally, interviewees noted that many students in the sciences were not strong writers, and that language models that assisted in writing would improve writing quality for a large number of students. P11 said, “I know there’s a lot of negative reactions to [ChatGPT] and professors are very concerned about [new forms of] plagiarism, and students are truncating the thinking process [by not writing]… but actually I think it [language models] will raise the floor in a lot of ways, for people who struggle with writing… there will still be room for great writers, [the model’s] not going to do all the work for you.” P19 noted that more than half of the graduate students in his department (engineering and geological sciences) were foreign-born students who spoke English as a second language; he noted that language models “will be a good thing in reducing inequities in being able to publish and write [for English language learners].”

\subsubsection{Education: Concerns about Generative AI}
Unsurprisingly, cheating was a common topic that participants brought up when reflecting on how Generative AI would impact education. While there was wide consensus among our participants that AI-assisted writing tools lowered the barrier to cheating (e.g., by producing written essays or answering homework questions) and made cheating harder to detect (i.e., due to the fluency of the produced writing), there was disagreement over the level of concern. 

Faculty who were less concerned about AI-assisted cheating tended to fall into three camps:

The first camp felt that (to paraphrase the singer Taylor Swift) “cheaters gonna cheat” – i.e., that students who wanted to cheat probably already did so, and that students not inclined to cheat were unlikely to be lured into cheating by the availability of large language models. Some faculty felt that Generative AI tools would be like other technologies that first created panic about education loss but eventually became accepted as valuable tools; P17 felt that AI would eventually come to be viewed “like a calculator,” though noted that (as with calculators) educators and students will need to reflect carefully on when and how to do tasks with and without the scaffolding the tool provides. Similarly, P9 emphasized that it will be important to teach students how to use Generative AI effectively, making an analogy to learning to use a search engine: “maybe it’s a tool you need to learn, 20 years ago I did not know how to use Google search… now I have the skill to search something I am looking for…” Indeed, P13 noted that it would be “hypocritical” for faculty not to allow students to use AI as an educational tool, given that scientists themselves saw value in how they could use AI as a tool to augment their own workflows. 

The second camp viewed the potential for AI-assisted cheating as an opportunity - a kind of wake-up call for faculty to overhaul evaluative methods to formats that might be more effective. P19 observed that an increased reliance on oral exams (already a common practice for many graduate-level students in the sciences) would be one way to change evaluation practices. 

The third camp noted that AI-assisted cheating might be a concern for introductory level undergraduate courses, but was much less likely to be an issue for advanced undergraduate coursework or for graduate students, since current large language models lacked the domain-specific knowledge to produce accurate answers for more advanced science courses. P10 suggested that perhaps someday, future AI tools might get to the knowledge level of a “third year graduate student,” but did not believe they would ever possess the highly specialized domain knowledge of more senior doctoral candidates (whose job is to create new knowledge). In addition to the lack of sufficient domain-specific knowledge to successfully cheat in advanced science courses, faculty also noted that the level of writing fluency was probably also not sufficient for cheating even in introductory undergraduate courses at elite universities where students’ standard of work is quite high; for instance, P20 noted that ChatGPT generates text on the level of a “C student” at an elite university. 

Though some interviewees were not overly concerned about cheating with AI tools, many participants subscribed to the view that AI-assisted cheating was problematic, and would allow cheating at a previously unseen scale (by lowering the barrier to cheating on a wide variety of assignment types - homework problem sets, essays, coding, etc.) and by making such cheating more difficult to detect. P8 observed that though in an ideal world faculty would spend time generating new types of assignments and exams that cannot be circumvented with AI, in the real world faculty are already overworked and that many “can’t afford [the time] to create … new exam problems.” AI-assisted cheating risks creating an unlevel playing field in science education and makes it challenging for faculty to evaluate students’ true understanding of concepts. Widespread use of Generative AI by students (or perceptions of widespread use) might reduce student motivation (since they will perceive assessments as inherently unfair due to high cheating rates) and may also reduce the prestige of scientific awards and venues (e.g., winning a graduate or postdoctoral fellowship from a particular scientific body might be viewed as a lesser honor if there is suspicion that fellowship proposals may have been written by AI).  

Another concern is that even if uses of Generative AI tools eventually become normative and are not technically “cheating” at coursework, students will nonetheless be “cheating” themselves out of key aspects of their education. For instance, P8 noted that the use of AI-assisted tools for summarizing ideas from course readings or for writing drafts of essays risks robbing students of opportunities for critical reflection, and that students with easy access to language models will lose motivation to spend time on the “creative process” of synthesizing concepts. 

Beyond concerns about cheating, some participants expressed concern that, even when using Generative AI in officially sanctioned ways such as for tutoring or curiosity-based Q\&A, students might be exposed to factuality errors. This was particularly a concern for more junior students (i.e., undergraduates), since people with less domain-specific knowledge might be less able to detect hallucinations, misinformation, or other types of errors produced by language models; P12 noted, “in the education part… people can totally get steered in the wrong direction… [if] someone they are learning from doesn’t have great knowledge.” In addition to outright falsehoods, another educational risk is that students might accept unquestioningly information from a language model without considering how biases in the model might be providing them with only one of several possible perspectives on a scientific question – P14 described this risk as the “politics of summarization,” nothing that what a model chooses to emphasize or exclude from a summary of a scientific topic could influence discourse. P2 notes that scientists themselves often disagree on facts, and that scientific understandings often evolve over time, emphasizing the nuance between whether an AI teaches aspiring scientists about a popular point of view versus about “how to move past the popular point of view.”

\subsection{Data}
Data are core to many aspects of the scientific method. Data may represent observations; analyzing these observations can support the generation of hypotheses. Experiments designed to test hypotheses can in turn generate data, analysis of which can support drawing conclusions and iterating on predictions, experiments, etc. Our participants saw many opportunities for Generative AI technologies to support data acquisition, preparation, and analysis, though they had concerns about the possibility of introducing errors into data sets, the removal of human reflection from data analysis, and the applicability of modern ML methods to “small data” scientific fields. 

\subsubsection{Data: Opportunities for Generative AI}
The opportunity for Generative AI to improve the efficiency of preparing and using data and/or to create new data sources was one of the most frequent themes in the interviews. The most commonly mentioned challenge faced by scientists in a wide range of fields was the need to do extensive labor in assembling datasets and preparing them for use. This includes tasks such as merging related data curated by disparate sources (e.g., satellite observations of weather data from multiple government agencies), which necessitates first tracking down and gaining access to each data source and then transforming the data into compatible formats (since each source often has its own metadata standards). In addition to merging tasks, data cleaning is a laborious and time-consuming task wherein incorrect, incomplete, or duplicate data must be removed. Finally, in many disciplines there are data labeling tasks that currently require time and expense to generate labels either from expert human labelers (e.g., speech pathologists labeling aspects of acoustic signals) or from paid crowd workers for less specialized tasks (e.g., transcribing the content of speech).  Graduate students may spend months or even years assembling and preparing data sets before they are able to proceed with scientific analyses of the data. 

For instance, P19 (a professor of geological sciences) noted that data in his field are generally public, but that “data being public doesn’t mean they are usable.” He identified the volume of data wrangling and data cleaning work that graduate students must do in his field as “the central bottleneck [to research].” The ability to issue queries in natural language to an AI model that would allow it to perform some or all of the steps involved in dataset preparation (or at least to generate code that would handle data processing steps) would be a huge boost to scientific efficiency. 

The volume and heterogeneity of data in many fields makes it challenging for many scientists to identify useful patterns. P20 (whose speciality is environmental remote sensing) noted that there is now more data than ever in her field due to the availability of satellite imagery. P9 (a climate scientist) said that several petabytes of climate data are generated every single day from sensor observations, simulations, and experiments. P18 (a bioinformatician) observed that in precision medicine there are so many time points and things being measured that a “data reduction” process is necessary for the scientists to limit what information they are able to consider in an analysis. AI systems that could identify patterns at scale in large, heterogeneous datasets could reveal new insights in many fields. Such systems could offer interfaces that allow scientists to query the data using natural language; for instance, P12 (who develops quantum computing hardware) noted that quantum hardware experts currently rely on “hunches” to debug malfunctioning hardware - they have access to data, but this data is not in a format that is useful for querying in order to make decisions because of its heterogeneous and distributed nature. 

Two of the social scientists (P13, a sociologist and P17, a medical anthropologist) noted the potential for Generative AI systems to expedite and standardize the process of qualitative data analysis. Analysis of qualitative data (e.g., thematic analysis, open coding, grounded theory, etc.) is typically a slow and labor-intensive process, requires extensive training to become expert in a particular method, and can be difficult to precisely describe and replicate. P13 sees the potential for language or multimodal models to “be a method in our [social science] data analysis toolkit,” but notes that there is a need for transparency and explainability of what the AI did and how for this to be accepted in the community: “we pride ourselves on our methodology… if we can’t share enough about our approach [to data analysis]... we can’t publish on it.” P17 emphasized that having more structured, automated approaches to qualitative data analysis can increase the extent to which scientific agencies make use of qualitative data (citing a recent interaction with the U.S. FDA who struggled to handle data of this type) – this can help ensure that qualitative analysis isn’t “demoted as a tool” relative to quantitative analysis. 

Generative AI models can also provide value in the assembly of entirely new datasets to support scientific inquiry. This could be through the generation of synthetic data; for example, P20 noted the opportunity to create synthetic image data to accelerate geosciences by modeling data that cannot be directly captured by satellites due to issues such as cloud cover or political restrictions on data collection (e.g., not being able to fly a scientific satellite over Russia or China). As an alternative to synthesizing data, AI could support dataset creation through the use of a language model to generate interactive prompts that support data collection from human subjects. For instance, P11 noted that language models could be used to create stimuli to gather structured speech data from end-users, and that such data might be more realistic than many current data-collection approaches if AI agents take a conversational approach to eliciting speech (rather than asking participants to simply repeat static speech prompts). Another approach to developing novel datasets using AI is to ask a language model trained on scientific corpora to crawl a set of papers and extract data that might currently be reported across a wide number of sources in varying formats and levels of detail, and transform it into a usable dataset. For example, P7 commented that “it’s a bit of a black art” to synthesize proteins in a lab (postdoctoral researchers must optimize several parameters such as what kind of cell to make the protein in, what kind of broth to put the cell in, at what temperature, for how long, etc.); extracting datasets of common methods from reputable research papers would then allow researchers to make natural-language queries to a model to get back recommended parameters for tasks of this type.

The need to compress high volumes of data available in some fields in order to host it within affordable infrastructure accessible to scientists at universities is also something potentially addressable by modern ML technologies – i.e., if a model were built of the relevant data, then simply storing the embedding-space representation (rather than the original data points), might be an effective compression technique. P6 suggested that embeddings might solve this “data transfer” problem so that scientists did not need to pay “half a million bucks” for storage, though he noted that “traditionalists” might object to such a representation since they would not be able to label each variable with a physical quantity (in his field of hydrology). P2 expressed a similar idea, noting that the machine learning model of a complex physical phenomena might be the most compact representation; he noted that such models would be inherently not understandable because they represent complex systems with trillions of parameters, but if the predictions of black-box computational scientific models are accurate, P2 felt it would be acceptable for scientists to use them even if they do not fully understand them.

\subsubsection{Data: Concerns about Generative AI}
Although participants were generally keen on the opportunities for Generative AI to enhance many aspects of scientific interactions with data, they did express some caveats to this enthusiasm. For instance, while P4 was excited about the potential for language models to extract unstructured information from scholarly articles in order to create new datasets, she noted that this data would need to be “trustworthy” and wondered whether using language models to process research articles for data extraction might introduce errors into the resultant data. Though P11 saw an opportunity for Generative AI to accelerate the slow and expensive process of data labeling, he also wondered about the accuracy with which AI could label complex data such as speech data that has “1000 different acoustic features” (though he also acknowledged that labels generated by human experts are “not 100\% accurate”). 

While P17 saw potential for language models to introduce standardization and efficiency into qualitative data analysis techniques, she noted that completely automating qualitative data analysis would remove aspects of human reflection that are considered fundamental to anthropological practice.  

P11 (a speech-language pathologist who works on tools for people with communication disorders) noted that “small data” was a challenge in his field (e.g., relatively few people have particular communication challenges, such as ALS), and wondered whether Generative AI models (which usually learn from large data sets, often with billions or even trillions of data parameters) would be applicable to his speciality. Similarly, P2 (a geophysicist working on earthquake forecasting) noted that because large earthquakes are rare events, there may not be enough data to train an ML model to predict them (and it may take decades or even centuries to verify whether such a prediction worked).

Finally, P2 expressed concern about Generative AI being used to create fake data, noting that fake data was a particular risk in fields like the life sciences, where data is often not shared openly (as compared to in earth science, physics, and astronomy, where open data sharing is the norm). Closed datasets would make fake data more difficult to detect, since it would not be inspectable by peers, and P2 feared that people will generate fake data (such as PCR data, genome data, or protein structure data) to attract funding for bioscience startup companies and/or to support fake “discoveries” in order to burnish academic reputations. 

\subsection{Literature Reviews}
Keeping abreast of the scientific literature in one’s field is a key part of scientists’ jobs. This is important for myriad reasons, including maintaining fresh professional skills through an ongoing awareness of key trends and discoveries, finding inspiration for future research ideas, identifying potential collaborators, and conducting literature reviews before embarking on new research projects in order to validate the novelty of a contribution, build appropriately on prior knowledge (such as by using comparable methods and data to past experiments for replicability and comparability), and appropriately citing related work as part of the scholarly publication process. Generative AI, particularly language models, are poised to substantially alter many aspects of the literature review process. 

\subsubsection{Literature Reviews: Opportunities for Generative AI}
As the pace and scale of science increases, it is increasingly challenging for scientists to keep abreast of the latest developments (and to sort the high quality articles worth reading from low quality articles or outright scientific spam). Participants saw great value in having an AI tool that could surface relevant articles to read based on a natural language description of a topic of interest (e.g., articles about a particular subfield, articles relevant to a specific hypothesis, articles using a particular methodology, etc.). P2 noted this could be a valuable tool for helping graduate students onboard to an area of study, noting, “this generation of students doesn’t devote their weekends to reading the old literature…” and suggesting that new ways to help surface relevant literature would benefit students. P3 echoed this idea of language models helping scientists come up to speed on a new topic, noting the high volume of literature scientists must digest when they embark upon a new project: “You just read paper after paper after paper…” Similarly, P6 noted that “the scientific literature is too large to be useful.” P18 noted shortcomings of status quo academic search engines for dealing with the high volume of literature in her field (stem cell biology and bioinformatics), giving an example of searching for a particular protein, P53 – she noted this would return potentially tens of millions of research articles (since P53 is involved in half of all cancers). 

While producing lists of related work to read in response to a query was desirable, participants such as P5 noted the importance that language models directly link suggested citations to the source articles, so that scientists could verify authenticity and directly inspect an article firsthand. Further, participants indicated that they wanted more than just a list of references, but would also value information about why each suggested reference was considered relevant to their query; for instance, P7 described wanting to know how a set of citations were interrelated, and P13 also desired explanations of why an AI would suggest a particular citation so that he could then decide whether it was worth his time to read the full article.  

In addition to surfacing (and summarizing and synthesizing) academic literature, P12 noted the opportunity for Generative AI tools to help scientists discover a range of relevant online resources, perhaps targeted toward their expertise or educational level. For example, a query on a topic from an undergraduate student might point to YouTube videos or slide decks, whereas a query from a graduate student might yield recordings of research seminars in addition to original articles. Multimodal content (figures from sides, clips from videos) could also be interspersed with text to create compelling, interactive summaries of areas for scientists of all abilities. P16, who works in quantum algorithms, noted that because quantum computing is such a young field, most knowledge about even introductory topics may require reading a “40 page research paper”; he emphasized the potential for AI tools to summarize concepts in more accessible ways for non-experts or new graduate students, reducing the “unusually high burden in terms of needing to parse the literature” in his discipline. 

P4 observed that in addition to identifying a set of articles to read, models that can suggest specific scientists working on particular topics (based on parsing the literature) would be another asset to graduate students or others coming up to speed in a new area. This can help scientists understand the key players and institutions in a sub-field, identify people whose work they want to follow, and even suggest partners for collaboration. P13 noted that junior academics in his field need to learn “who are the groups of sociologists that work together, publish together, think together that you would want to be a part of.” He also noted that literature review tools could help junior scholars identify relevant venues for publishing their own work by giving insight into who publishes in certain journals, who cites work from those journals, etc. 

P6 commented that in science there are often “silos” of researchers who cite each others’ work (sometimes influenced by geography), and that AI that could identify scientists conducting relevant, high-quality work from outside one’s typical citation graph could bridge such silos, potentially resulting in more scientific advances. P13 noted that in addition to making scientists aware of work outside of their typical citation networks, such tools might actively promote DEI agendas, such as helping give more visibility to scholars from historically marginalized demographics (e.g., women, Black, Latinx, etc.), from a broader set of institutions (e.g., HBCUs), and/or from the Global South. Further, P13 noted that many journal articles are inaccessible to scientists from institutions with fewer financial resources due to paywalls, and that AI tools that can summarize such literature could broaden access to scientific knowledge (though circumventing paywalls through AI summarization would introduce its own set of ethical and IP issues). 

Several participants noted the potential for LLMs to add value to a literature review by not merely recommending articles but actually synthesizing summaries of a topic to accelerate learning about a new area, going beyond search engine functionality to actually “assembling knowledge in usable form” as P1 described; he envisioned an intelligent assistant that would “read articles and summarize them” for the scientist, so that they could then determine which articles to invest in reading fully. Similarly, P7 wished for a tool that could succinctly summarize the key points of any particular academic article. 

Both P3 and P4 noted a desire for tools that would help synthesize and surface insights from a collection of articles; P3 noted that such connections might “be missed just due to bandwidth limitations of scientists [conducting a large literature review].” While P1’s vision was more around short summaries of individual articles, P3’s vision was more around compression of a body of scientific literature (he described the medical literature as “massive”); he envisioned he could search PubMed and request the top 200 search results on a topic and then ask an AI to “give me a two page digest of what all these [200] articles say.” 

P3 noted an opportunity not only for an AI to summarize articles, but to produce a customized review article on demand. Review articles appear in scholarly journals in many fields, and summarize and synthesize the state of the art by considering a large body of work. However, as P4 noted, review articles can quickly become out of date (sometimes even by the time they are published, due to the quick pace of many scientific fields combined with the slow pace of journal publication processes). Further, review articles simply may not exist for some topics, as P3 noted that when coming up to speed on literature in preparation for a new project, “you read review papers as well… [but] sometimes those kinds of review articles don’t exist.” P12 also expressed a desire to get a set of references related to a “niche topic,” unlikely to be covered by existing digests. P13 noted that sociologists “can’t read everything [in their field]” and specifically depend on “annual review pieces” in which a famous senior researcher will “write a state of the union on [their topic of study]... and we ALL read that.” Book reviews are also particularly important in sociology; P13 described how famous sociologists would read each others' books and write review pieces, which were highly valued as reflections by others in the field. AI tools that could produce annual review pieces and book reviews of this type could provide value to other sociologists, though it remains to be seen whether the timeliness and specialization of AI-authored review pieces would outweigh the desire of scientists to get such summaries distilled through the unique perspective of particular senior members of their profession.  

Language models to support literature review could not only summarize individual papers or synthesize findings from a set of papers, but could also find patterns in a field over time, revealing high-level trends that may not be apparent to newcomers (or even more established scientists) in an area. P15 expressed a desire for AI tools that could look over a body of new work and tell him “what’s hot” in a particular discipline. Further, such large-scale analyses could potentially identify “gaps in the literature” (P7), which could indicate opportunities for future exploration. A related concept was proposed by P18, who wanted to be able to ask an AI “what is already known about this space?” in order to save time by avoiding exploring research problems already well-addressed in the literature. 

In addition to suggesting topically relevant papers to scientists based on natural language descriptions of interests, P4 noted that it would be valuable if an AI system could learn from past patterns in the literature in order to predict the likely future impact of newly published papers (i.e., which papers are likely to be highly cited in the future, to win awards, etc.); this could help with the challenge of keeping abreast of a growing literature by directing scientists’ attention to items with a high probability of importance. P8 noted an explosion of low-quality scientific literature that added to the challenge of keeping up with relevant findings: “there are so many journals that don’t care about the quality for publications”; she suggested that AI tools could help identify “premier work” to direct scientists’ limited attentional resources.  

Staying abreast of the current literature in one’s own subfield of scientific expertise is quite challenging, as our participants described. The challenge of keeping up with their own field precludes most scientists from becoming aware of relevant discoveries and ideas from other scientific disciplines. However, interdisciplinary connections can be invaluable to scientific discovery. For instance, P3 notes that such connections can sometimes lead to a “quantum advance” in science, noting how concepts from physics substantially advanced areas of computational biology. AI tools that could recognize relevant patterns or parallels across fields and proactively suggest relevant literature across disciplines could facilitate serendipitous discoveries. Similarly, P15 noted that their university department specifically sought to hire faculty working at the intersection of multiple fields because they viewed these intersections as areas with great potential for scientific advancement; he described the current process used to figure out if methods or tools from one domain apply to another as a bit of a “random walk process” and suggested that an AI that is scanning everything in the scientific community and can recognize potential connections for how an advance in one field could be applied to another would add tremendous value. P5 (a materials scientist) saw value in new methods of querying for related work across disciplines, such as by asking an AI in natural language about a set of materials, and then getting back a cluster of research areas that would be relevant (e.g., learning that for a certain combination of materials that “solar cell researchers are interested in these and also biomedical researchers are,” which might yield insights into aspects of those materials that had not previously been considered). P14 noted that currently, conducting interdisciplinary literature searches can be challenging because of the need to figure out key words used in each field, “there may be the same concept that is called something different across [fields]”; AI tools that could bridge this vocabulary problem could also accelerate interdisciplinary work. 

Several participants expressed a desire for language models to help them write (or at least draft) literature review sections of scientific papers. For instance, P6 envisioned being able to enter a research topic into an AI tool and get out a fully written literature review or related work section. P9 hoped AI tools could automate writing literature summaries required by many grant applications. We discuss this possibility more in the section on AI-assisted scientific communication. 

Language models could also provide interactive reading-support tools for use while reading a particular article. For example, P16 notes that if a paper mentions a particular concept (or jargon-y term) that is unfamiliar, a student could ask the model to explain the term to them in order to support their comprehension of the article (without requiring an additional literature search).  

In addition to conducting literature reviews to learn more about a new topic or stay abreast of one’s current field of expertise, our participants also described conducting focused literature reviews to manually extract unstructured data from large sets of research papers, such as precise details of experimental methods or simulation parameters. For instance, P7 described reading a series of papers to identify optimal conditions other scientists used to synthesize a particular protein in the lab as a precursor to designing her own experiment. She described another time when she conducted an extensive literature search to inform her experimental design, where she had been “just manually creating spreadsheets of citations from the literature of what sort of experiment did they [other scientists] do, what was the finding… for over 100 different papers, pulling out these exact numbers [a particular measure she was interested in].” This process of structured data extraction can be extremely time consuming, and could be automated by AI tools. 

\subsubsection{Literature Reviews: Concerns about Generative AI}
Hallucinations were a major concern for using Generative AI in learning about the scientific literature. P2 noted that he tried using Meta’s Galactica AI \cite{taylor2022galactica} (released briefly in 2022 and billed as a “language model for science”) to ask basic questions about earth science, and described the responses as “horrible,” sharing incorrect information such as that “earthquakes are the most powerful force in the solar system” or stating incorrectly that Mt. Everest is made of igneous rock. P11 noted that the fact that status quo LLMs such as ChatGPT don’t provide links to references is “a huge limitation… as an academic, you just can’t take it [ChatGPT] for its word,” noting that the need to track down and verify references manually would reduce any potential time savings of LLM-powered literature searches. P13 also expressed reservations about hallucination and “misinformation” in AI-assisted literature reviews; she noted that because current interfaces to LLMs are similar to those of search engines (a simple textbox), that end-users might have a mental model that they are simply receiving search results akin to using a tool like Google Scholar (which might lead to a level of trust inappropriate for generative AI outputs). Factuality remains a barrier to successful adoption of language models for literature searches, until scientists can trust that models will not hallucinate scientific facts when summarizing papers or invent citations entirely. For example, Galactica was also found to invent plausible sounding paper titles, authors, and journals that did not exist, sometimes mixing and matching real and hallucinated content such as a real scientist’s name as the author of a non-existent article. P20 relayed an experience where a colleague of hers had asked ChatGPT for a set of references on a topic, and the tool returned a nicely formatted reference list with the names of real scientists in her field and real journals in her field, but completely fake paper names.

P14 noted that while summaries or syntheses of individual or sets of scholarly articles could accelerate research, such summaries inherently introduce bias in the choice of what is included or excluded from a synopsis; she described LLM-based scientific summarization as “a critical editorial moment,” and noted that many scientists might rely on such summaries and never check the original source material, thereby remaining unaware not only of any outright inaccuracies in summaries but also of the “politics of summarization,” i.e., decisions made by the AI that might skew interpretation. Another bias risk is that rather than breaking down citation silos, AI systems might reinforce them by systematically directing attention to particular works at the cost of other articles. 

P5 wondered whether Generative AI was truly necessary to support keeping abreast of literature; in order to adopt AI tools for this task over status quo methods (Google and Google Scholar searches), there would need to be clear metrics proving that the use of new AI models improved either efficiency or comprehensiveness of the literature search process. These improvements in efficiency and/or comprehensiveness would also need to outweigh the possible loss of critical insights gained from the process of reading, reflecting on, and synthesizing literature search results. P12 reflected on concerns about “how [AI] shapes our way of knowing… and what is lost through those summarizations.” However, because literature searches are an area in which many scientists already use technology (search engines), there may be less social resistance to experimenting with novel AI systems for this purpose than for other aspects of scientific work.

\subsection{Coding}
Generative AI systems such as OpenAI’s Codex and GitHub’s Copilot \cite{chen2021evaluating}, which debuted in 2021, are specifically designed to complement (or potentially replace) human programmers by producing software snippets that can function as-is or that can be edited by a human developer to accelerate the coding process. In addition to coding-specific AI tools, many general-purpose LLMs can also produce code as an emergent behavior \cite{sparksofagi}, since code samples are present in many of the internet forums used for training models. Computation is increasingly a part of many fields of science; programming is necessary at various stages of scientific work, including for collecting or cleaning data, for running simulations or other experimental procedures, and for running statistical analysis on experimental results. The majority of scientists are not formally trained in computer science or software development, and producing correct code (much less high-quality code that is efficient and reusable) is extremely time-consuming for this constituency. AI tools that can accelerate or automate coding via a natural language interface have a huge potential to accelerate scientific discovery. 

\subsubsection{Coding: Opportunities for Generative AI}
Several participants mentioned that they have already tried using status quo LLMs (such as ChatGPT) to make natural language requests for code. P16 (who designs quantum algorithms) notes that he asked ChatGPT to compose a quantum algorithm and “it is able to do that somewhat… it’s not particularly good at it, but it’s better than I would have expected.” P11 also noted having tried coding with ChatGPT, observing that “it wasn’t great… it just made up code that looked good but didn’t work”; however, P11 pointed out that if future AI could generate accurate code it would be “game changing,” estimating it would make him “at least 30 - 40\% faster” at coding, which he noted occupied a substantial portion of his time doing speech-language pathology research. 

P18 (a bioinformatician) described the extent to which coding skills are intertwined with her field, noting that every research lab rolled their own custom research software, a time-consuming process: “Right now I think the biggest problem [in bioinformatics] is that everybody is working on these systems that are self-generated, every lab has set up their own servers, cloud service, pipelines… setting up your environment to even run packages that are already established can be months…” Further, P18 noted that many labs had to re-do this process every few years as students and postdocs moved on (since the code tends to be so brittle and poorly documented that it could no longer be run when the original authors left): “when somebody leaves a lab you’ve lost institutional knowledge [about code].” Similarly, P10 (a quantum chemist) noted that a challenge for first year graduate students in his field was the many types of input files they needed to code in order to use established quantum chemistry packages; he further noted that this learning process was compounded by the poor documentation of status quo quantum chemistry code. P9 (a climate scientist) similarly noted problems with the quality of code written by people without a software engineering background, observing that poorly designed code in climate models (often comprising millions of lines of FORTRAN) resulted in other scientists having difficulty using the models. Generative AI tools that not only write code, but also write code that is well-structured and well-documented in a manner that enables code learning and re-use, would eliminate a great deal of time and effort common in the sciences today. 

More structured and well-documented code would not only support code reuse within research labs, but would also support better code-sharing across institutions. P5 noted that a lack of programming standards and code-sharing across labs is currently a limiting factor for accelerating work in the physical sciences. 

Not only do few scientists in domains outside of computer science possess strong coding skills, but the need for coding skills (no matter how amateur) is itself increasingly a barrier to participation in the sciences. For instance, P2, P6, and P19 noted that lack of coding experience is a barrier to entry for students interested in getting into climate science. AI-assisted coding could broaden participation in the sciences, particularly from historically under-represented groups who are less likely to study programming (e.g., people identifying as women, Black, Latinx, etc.). 

Even if a Generative AI is not able to produce fully functional code, scientists saw value in tools that would accelerate their own coding by suggesting relevant APIs or syntax. For instance, P11 (who had tried coding using ChatGPT) noted that while he wasn’t thrilled with the quality of generated code, he found it a useful tool for learning about available package functionality: “it was faster than me looking it up” and that he found AI helpful for answering questions like “how do you use this function.” P11 summarized the value-add of an AI that could explain APIs: “for someone like myself, I’m not a full-time programmer, so I do need to reference things a lot.” AI support for referencing API information would also benefit experienced programmers who must learn to deal with legacy code systems – for instance, several participants mentioned that legacy scientific systems in fields such as physics and climate science are written in FORTRAN, a language not typically taught in modern computer science courses; AI tools that could allow scientists to quickly come up to speed in unfamiliar programming languages could be of great benefit.  

P16 also noted that AI-assisted coding could be a timesaver for students doing homework problems, since students currently have to learn complex APIs for running simulations or other types of exercises, where the writing of the code is not the pedagogical purpose of the assignment. 

Writing code to do statistical analyses of data (e.g., scripts for packages such as R), is another area where Generative AI could accelerate science by not only automating or accelerating the scripting itself, but also by suggesting appropriate statistical tests, since many scientists are not experts in statistical analysis. P18 noted that one benefit to an AI tool would be if one could “put on some guard rails so people do not do inappropriate things with statistics.” P14 noted that statistical software packages such as R and Stata could be “burdensome” to use, and that using natural language to request statistical processing code would improve their workflow. P3 noted that many biologists are not skilled at doing statistical power calculations (and that doing these incorrectly often leads to inconclusive experimental designs); tools that support natural language specification of statistical tasks could therefore support experimental design as well as analysis. 

\subsubsection{Coding: Concerns about Generative AI}
The main concerns participants expressed around coding were around correctness; if generated code does not work correctly (particularly if bugs are subtle), it may not save effort if scientists have to spend time teasing out the errors (or may cause scientific errors if scientists are over trusting of buggy AI-generated code). IP was an additional concern, i.e., if an AI tool proposed code that was proprietary to a third party (due to a “memorization” error), scientists wondered if this could lead the original authors of the code to have an ownership claim to their discoveries.; for instance, P19 worried that it was “risky” to use AI-generated code, as others might claim credit for subsequent work if the model had trained on their sourcecode.   

\subsection{Discovery}
Generative AI has the potential to directly accelerate the discovery of new scientific knowledge through improved experimental design, novel methods and models, and support for interpretation of experimental results. In this section, we share scientists’ reflections on how new AI models can support various aspects of scientific discovery, as well as cautions to be wary of.

\subsubsection{Discovery: Opportunities for Generative AI}
Several participants noted opportunities for Generative AI models to help scientists better direct their attention and resources, such as by helping them choose which of several hypotheses might be most likely to yield results, what data to gather to support their inquiry, or how to optimize experimental protocols and conditions. For instance, P2 noted that the placement of monitoring equipment is critical for studying earthquakes - AI models could help earthquake scientists plan the optimal layout of their sensor networks in order to maximize the value of the collected data. P6 noted that AI models could help model the ideal places for field hydrologists to take measurements (today measurements are often done based on areas that are easy to access rather than those that might theoretically yield the most information). Similarly, P8 expressed an interest in AI-powered tools that would help her determine how to best invest resources by predicting the probability of success of a particular hypothesis. P10, a quantum chemist, saw value in AI tools that could suggest to him what methods to try next if a particular approach to a problem fails, helping to quantify the probability of which of several computational or theoretical approaches would be most suitable. 

P6 went further, speculating on whether AI models several generations more advanced than the status quo might be able to not just help scientists refine hypotheses, but actually generate new directions for exploration. He commented, “the process of generating a question to ask is the hard part of science… and in fifty years I can easily see AI changing this part of the process.” However, this view was uncommon - most participants expressed that they did not think technology could ever replace the role of human scientists in having the creative insight to generate novel research ideas. For instance, P12 felt that it would be hard for Generative AI tools to propose novel scientific research questions since, “there’s so much context that would be extremely challenging to give some type of ML system… that kind of lives in people’s heads, that’s not in any type of tangible form you could give to a computer.” P19 observed that by definition, research is supposed to “extrapolating from what’s there,” which indicates to him that by definition models trained only on past data could never generate novel research ideas. P3 summed up this sentiment about the limited role AI was likely to play in identifying novel research questions by noting that scientists typically must first understand the “bleeding edge” of a field in order to choose “what are the questions worth answering.”

P3 (an immunologist) noted the complexities of experimental design, particularly in designing an experiment that would yield definitive results (versus one in which several alternate hypotheses might plausibly describe the findings); he also noted that the optimal design is not always possible (i.e., the experiment might cost too much, or require too long a time period) – an AI system that could provide advice or feedback on experimental designs (including within practical constraints), or one that could help a scientist select which of several potential designs would be most likely to yield insights could be a valuable tool. In the life sciences, P3 notes that there are startups such as Emerald Cloud Lab that have been developing platforms to help automate certain kinds of experimental designs, but that using such platforms requires creating regimented schema – the ability to interact in natural language with a LLM would make such approaches much more intuitive and broadly adoptable than relying on formal specification languages. 

P4 notes that in materials science, “experimentalist” approaches to materials discovery typically involve a substantial trial and error component – AI tools that can consider a broad range of experimental designs and suggest the most promising ones to actually try in the lab based on analyses of prior literature in the field could greatly accelerate materials discovery. P16 also wished for tools that could be used for “predicting the effectiveness of an experimental design,” where he could ask questions like, “what is the most effective way of mitigating errors in the measurements of this experiment,” and perhaps the model could even compose code that would be used to run the experiment (i.e., in a simulation environment). P18 noted that doing physical (lab) experiments in the biosciences is a “major investment” - that having confidence in the likely utility of one experimental design over another would help scientists have “prioritized targets” on how to direct their attention. 

P7 identified an opportunity for AI tools to support experimental design in biochemistry by helping graduate students plan details such as how much of various ingredients they need to make (e.g., a certain amount of buffer with a particular concentration). Currently, many students manually create complex spreadsheets to plan such details; computational tools that could take natural language inputs and knowledge of methods culled from the research literature could greatly simplify this process. 

Beyond lab-based experiments, P5 discussed how Generative AI could potentially support simulation-based work, such as by  helping scientists choose which simulation settings would be most appropriate to investigate a particular research question. P5 further suggested that AI could increase the efficiency of simulations by predicting some properties (e.g., properties of materials in materials science) so that only partial simulations are required. 

P11 described how Generative AI could be useful in his field (speech language pathology) by enabling new research methods. For instance, he could ask a language model to generate a set of speech stimuli that meet particular criteria (e.g., a set of ten sentences at least fifteen words in length). AI chatbots that could engage in conversations with patients could not only be a novel therapeutic tool, but could also be a method of collecting data about patients’ changing abilities, as well as serving as a mechanism to provide feedback to patients about their speech output. 

P1 (a behavioral economist) observed that language models could potentially replace human participants in some types of social science research; P1 gave the example of a paper he read recently where a fellow economist used a status quo LLM to replace the human participants in the classic “ultimatum game” experiment \cite{aher2023using}. This approach could be useful for conducting new studies (though the correlation between modeled participants and human subjects remains to be shown, but is plausible, just as participants from online crowd platforms have increasingly replaced in-lab participants for many social science studies in the past decade); alternatively, studies with simulated participants could be used as a low-cost method to test the replicability of social science findings, which could lend confidence to fields such as psychology that have experienced a replicability crisis.  

Generative AI models can potentially provide a new source of insight for social scientists studying societal-scale phenomena, providing a new medium for study. For instance, Google Search famously created opportunities for “nowcasting” of public health trends such as the spread of the flu (e.g., Google Flu Trends \cite{Ginsberg2009}) or macroeconomic trends (e.g., the volume of people searching for the address of unemployment offices as a leading indicator of unemployment numbers). The social scientists in our study were all interested in how they could utilize language models or multimodal models to gain insight into societal trends in fields such as economics, anthropology, and sociology. It remains unclear to what degree an end-user of a model can use it to gain insights versus whether using models to obtain social science insights will require “back-end” access to developer metrics (e.g., the frequency of certain prompts or responses); the stochastic nature of most Generative AI tools may also introduce replicability challenges in using them as a material for study and experimentation. 

As discussed in the Data section, AI tools that can analyze large, heterogeneous data sets can be used to model complex phenomena and to discover patterns that are at a scale impractical for human analysis. For instance, P2  notes the potential to predict earthquakes (without physics-based models) if it were possible to combine a huge set of heterogeneous observations about various phenomena (e.g., satellite data, GPS data, weather data, etc.). Similarly, P3 notes that precision medicine could be advanced if AI tools can help scientists discover patterns in huge data sets combining demographic information, behavioral data, medical history, and more.

\subsubsection{Discovery: Concerns about Generative AI}
P5 expressed concerns that Generative AI could exaggerate “fads” in scientific research. For instance, in materials science, he noted that many researchers began exploring properties of graphene after it was first discovered, leading to a glut of research on that topic and an opportunity cost in not exploring other materials. If many scientists use machine learning models to choose what to explore next, this might exaggerate such “fads”; while this may not be harmful at a micro-level, at a macro-level it could slow down progress in a field. P19 indicated that he suspected AI would “generate fads of large groups of people [scientists] using what’s available…” (i.e., choosing problems or approaches the AI tools make easy).

Accuracy and factuality remain key concerns for using AI in the discovery process. For example, while a conversation with an LLM-powered chatbot could potentially be a future diagnostic method for various disorders that impact speech or cognition, the risk of “misdiagnosis without human validation” is of concern to P11, who notes the importance of understanding the costs of false positives and false negatives for different use cases of AI, and ensuring that “we get human values into this process.” On the other hand, P15 pointed out that, while bad science fueled by AI errors is clearly not ideal, bad science already happens - there are huge numbers of poorly designed or interpreted experiments, non-replicable studies, studies published in low-quality journals, etc.; scientific findings fueled by novel AI would need to be verified and replicated in the same way as scientific findings fueled by status quo methods. However, the existing volume of low-quality science could reduce the efficacy of any generative models trained on that literature, and risk amplifying ineffective methods or avenues of exploration; as P18 noted, “large amounts of scientific literature are not reproducible… there is a lot of noise in that data.”

Several participants expressed concerns that Generative AI, by its very nature, would not be useful for the discovery of new knowledge, since such systems are trained only on data from the past. If such systems can only extrapolate from past patterns, it is not clear a priori that they can ever predict new concepts. If accurate, this point of view suggests that having a vision or hypotheses will remain the unique role of scientists, even in an AI-enhanced future. For instance, P5 observed that widespread use of AI tools in the sciences might make research overall less novel and more incremental, because models might be likely to focus more on predictions that are highly related to past data rather than on more novel breakthroughs. P6 and P9 also questioned whether generative models trained on past scientific data would generalize well to predicting future trends, particularly in fields impacted by climate change, where, for instance, future weather patterns are expected to differ quite substantially from the past. 

Finally, it is worth noting that if Generative AI can vastly accelerate scientific discovery, that might include not only socially beneficial discoveries (e.g., new medical treatments, clean energy sources, etc.) but also potentially malicious discoveries (e.g., biological or chemical weapons). Ensuring that scientific AI tools include safeguards against unethical applications is a key concern.

\subsection{Communication}
One of the biggest near-term opportunities for generative models is to support scientific communication, including the writing of scholarly articles, grant applications, and preparation of presentation materials such as slide decks. There is also an opportunity for AI tools to play the role of reviewer, offering preliminary feedback that helps scholars fine-tune their arguments ahead of true peer review. LLM-assisted scientific writing also carries many risks, particularly the risk of generating publication spam and scientific misinformation. 

\subsubsection{Communication: Opportunities for Generative AI}
The most common way in which interviewees envisioned using language models was to have the models produce “first drafts” of writing, which they would then edit. Sections of scholarly articles such as Related Work were particularly seen as ripe for AI-drafting, since tracking down citations was viewed by many as a time-consuming and tedious process (an idea which follows naturally from the discussions covered in the section on Literature Reviews). The need for an expert to validate the output of the writing, even for relatively rote sections such as Related Work, was on participants’ minds; P3 noted it would be most useful “if it can cite references that you can then go and check up on and make sure they actually support the statements.” P8 felt that the first draft of the Introduction section of many scientific articles could also be delegated to language models, which could help articulate the background and motivation for a particular research problem. P9 summed up this consensus view that LLMs would be useful for drafting (but not automating) scientific article writing, by noting his skepticism of quality scholarly writing with “a push of a button,” but nonetheless appreciated that an AI first-draft would help them “be more efficient.”

Alternatively, rather than asking a model to produce a first draft that would then be edited by the scientist, the scientist could provide a very rough draft (i.e., an outline or set of notes) to the model, which could then produce polished, grammatically correct and well-organized formal writing. P16 reflected on the utility of this approach: “the reality is a lot of scientists aren’t very good writers.” 

While writing journal and conference articles was considered to be core to scientists’ intellectual roles (and thus something that would benefit from partial but not complete automation), participants were more eager for tools that would automate grant writing. Many scientists (particularly those employed primarily at universities rather than primarily in industry) spend a great deal of time writing research grant applications, and tools to create comprehensive first drafts of grant proposals were highly desired. P19 observed that much of the text put into reports for funding agencies tended to be “fairly generic,” and thus ripe for AI automation. P20 viewed the potential for AI-assisted grant writing tools from a slightly different angle, observing that such tools could be educational for junior scientists, helping them learn about the best ways to phrase their ideas to be most successful in obtaining research grants. 

While participants unanimously agreed they would not trust a language model to autonomously compose a scholarly article on their behalf (nor would they necessarily want this even if they trusted a model were capable thereof, since the process of writing entails valuable critical reflection), they did see opportunities for future generations of LLMs to autonomously compose lower-stakes materials. For instance, P2 was eager for a generative AI that could relieve administrative bureaucracy load, such as responding to many of his emails. P19 also noted that much of his time at the university was spent on “humdrum boring stuff… emails, many of which have generic answers that I’m sure an automated system could generate easily”; he noted that while writing a scientific paper has a creative element that requires a human expert, a lot of the “office stuff” that scientists do could be delegated to language models. P1 observed that some simplified synopses of scientific findings aimed toward a lay audience could likely be fully automated (in the case of economics, for instance, he noted that stock market news reported in the media is quite formulaic). 

Participants were also eager for models that could reduce time spent on formatting tasks. For instance, P7 mentioned experimenting with ChatGPT, giving it a 400 word abstract she had written for a journal article and asking it to condense the abstract into 200 words to meet the length limits; she felt it did a pretty good job at that task; condensing text to meet word limits and fidgeting with formatting to meet the varying requirements of different submission venues is a tedious task that scientists would be happy to automate (and a task forseen by the human-in-the-loop system Soylent more than a decade ago \cite{soylent}). P14 emphasized the tedium of formatting citations (different submission venues require different citation conventions), and lamented that many scientists spend hours doing tedious and “seemingly pointless style editing” that could be automated (for instance by asking a language model to output LaTeX markup). 

Beyond language models, image or multimodal models can also play a role in time-savings for scientific communication. The ability to generate scientific figures (tables, charts, graphs, diagrams, etc.) using a natural language request was something identified by participants as useful both for the composition of scholarly articles as well as for presentation and lecture slides. “Something that can make figures would be powerful,” noted P3, who lamented how time consuming making figures can be, particularly in situations where there is a large amount of data that requires the scientist to figure out abstract ways to represent the data in order to communicate insights. P12 wanted to be able to interact with a model with a multimodal interaction, by saying “draw me a figure that looks like this,” and then giving an example of an image of a certain type (e.g., a rough sketch or an image from another article). P14 envisioned creating tables by providing the model with her data and then commanding it to “create a table with demographics… and I want it in [a particular] format.”

The ability to generate a slide presentation including text and figures from a set of inputs would also be highly valuable. This could be a presentation for a scholarly audience (taking a journal article and/or data as an input to the model), or it could be an educational lecture for students (taking a lesson plan or set of diverse materials as input and producing audience-appropriate slides). 

The ability to take a document (whether an article, slides, etc.) and have them adapted to be more suitable to a particular audience would also be a valued and time-saving application of language models – for instance, asking the model to tune a presentation or article for an audience with a particular background (scientists, the general public, students of a certain level) or for the constraints of a particular venue (i.e., a presentation with a particular length limit, producing bullets or summaries in formats required by particular grant agencies or journals, etc.). For instance, P4 noted that she might want to ask a model to recalibrate the level of esoteric “jargon” in a document depending on the intended audience. P16 envisioned that he could “give one of these models a paragraph, like here’s a very precise mathematical statement, can you make this a bit more understandable”; he noted this might be particularly valuable in an interdisciplinary field like quantum computing, in order to tailor writing for presentation to a particular audience who each prefer different terminology and levels of technical detail (i.e., mathematicians vs. computer scientists vs. physicists). 

P8 noted an opportunity for AI to assist in the process of reviewing scientific literature (i.e., to determine acceptance to journals or other publication venues). One possibility is that language models (particularly those trained on prior literature in the field) could do an initial review of a paper (i.e., checking for novelty with respect to prior work) – this could be a filter to determine whether a paper should make it through to a set of human reviewers, thus reducing reviewer workload. Language models could also lighten the workload of human reviewers by providing writing support, such as producing a grammatically correct review from a set of notes the reviewer has made about a submission or helping the reviewer track down relevant references. P14 also noted the growing challenge of managing reviewer workload in her field: “the publishing process was already strained before the pandemic… it got even harder over the pandemic, people were not available to review”; she lamented that limited reviewer time relative to submissions volume results in slowing down the dissemination of findings. AI systems that reduce reviewing workload could help address this problem (though there is also the risk that AI systems that lower the barrier to writing journal articles will amplify the problem by dramatically increasing submission rates). 

Alternatively, models could provide reviewer-like feedback about an article draft to an author as part of the writing process, helping them improve the article before they submit it for peer review. For instance, the model could ask them questions about the article, or even attempt to automatically write a review, in order to help the author proactively anticipate how to present their ideas more clearly and convincingly.

\subsubsection{Communication: Concerns about Generative AI}
P7 was particularly concerned about the use of language models to generate fake scientific publications that are actively intended to create disinformation about controversial (politicized) scientific topics, such as vaccine efficacy, gene editing, or climate change. She notes the danger of losing public trust in the authenticity of scientific articles: “science does have this important role in society of being able to give confidence when it’s right or wrong.”

Beyond active misinformation attempts, there was also concern about publication spam (i.e., the use of language models to flood the scientific ecosystem with either fake or low-quality papers meant to burnish an author’s CV); as P16 quipped, “AI tools may make it even easier and more efficient to do relatively low quality science.” P14 observed that publication spam could overwhelm the peer review process, and P15 lamented that publication spam would further exacerbate the challenge of deciding what papers to read to stay abreast of one’s field. P16 echoed this concern, that a proliferation of AI-authored low-quality articles would make it more difficult to “find the really valuable results that you need to be reading.” There is also a cyclical risk of publication spam poisoning next-generation AI models, if the newer models were to train on the AI-generated false or low-confidence information.

\subsection{Trust}
The issue of trusting AI was core to the interviews; participants frequently shared concerns relating to trust (e.g., concerns about hallucinations and factuality, a need for citation back to sources), and were also asked at the close of the interview to specifically reflect on what would be necessary to trust the use of Generative AI in the sciences. Addressing these concerns (some technical, some social) are vital for the development of AI that aligns with scientists’ and society’s values, so as not to risk diminishing trust in science itself. 

Citation (which is something scientists themselves are trained to do in their own writing) is of paramount importance. Participants wanted to easily be able to trace back to original source materials, not only for straightforward tasks like AI-assisted literature searches, but for the full range of tasks a scientific AI might support, such as being able to understand why the AI recommended a particular method or experimental design, what evidence supported its hypotheses, etc. Not only were citations necessary, but these citations should be as specific as possible (i.e., referencing the specific portion of an article from which a piece of information is drawn), and the citations should be highly trusted and of good quality (i.e., the AI should preferentially ground its suggestions in science from a particular set of trustworthy scholars, institutions, or publication venues). 

Factuality and hallucinations were a major concern. Participants who described interacting with status quo Generative AI tools (Meta’s Galactica, OpenAI’s ChatGPT, Google’s Bard) reported a lack of factuality in answers to scientific questions and in invention of nonexistent scientific references, as well as deviations from specified instructions (e.g., P11 asked an LLM to create a set of fifteen-word sentence prompts for speech pathology studies and the system returned prompts that had variable lengths such as eleven words). P2 also noted that the definition of factuality changes over time, as scientific knowledge progresses, and that if models are not updated frequently they may inaccurately reflect the current scientific consensus. Indeed, because models are trained on past data, they may be biased toward older scientific perspectives; P6 quipped that “ChatGPT is a Physical Hydrologist” based on his perception that its answers to hydrology questions reflected more traditional work in his field that is less representative of the most current trends. 

Confidence metrics were seen as a vital part of helping scientists gauge the level of trust to put in model outputs. For instance, P7 reflected on how she had tried using Meta’s Galactica science model, but was disturbed by the lack of confidence metrics on its outputs (which were entirely hallucinated for the questions she asked about common proteins).  The ability of an AI system to provide a top-N set of responses (with associated confidence metrics) rather than just a top-1 response could also help engender trust in the system. P12 had the related suggestion that he would like the AI to specifically inform him when it is not confident of something, noting that when human scientists are not confident in something they might offer an answer or explanation but then provide the caveat that it would be good to follow up with some additional references or a second opinion from another expert. 

Explainability was important to some; scientists wanted to be able understand why any scientific AI would make particular recommendations or predictions. P1 said, “I might like to be able to have it explain its answers,” noting that such explanations would be particularly crucial if the AI suggested something that humans viewed as unlikely. On the other hand, P2 indicated that in some places explainability may not be possible, speculating that perhaps embedding-space representations of complex phenomena that could never be understood by a person might be the most efficient models in certain domains, and that if such black-box models resulted in accurate predictions of physical phenomena, scientists should use them despite their opacity. 

Interactivity (including interactive explainability tools) might help engender trust. P6 described how “interactive dashboards” in which users could change model parameters and see the resulting outcomes helped him persuade water managers of the accuracy and utility of hydrology models (powered by more traditional ML systems). Another form of interactivity suggested by P12 would be having the AI provide feedback when it is given a task, to ensure that it is correctly interpreting the scientist’s instructions, since small nuances and misinterpretations in natural language interactions might dramatically alter outputs. 

Some participants mentioned a need for protocols to disclose whether an AI was involved in some aspect of the scientific process, so that other scientists could then perhaps view such work with a more skeptical eye. This might require the development of new tools (e.g., to determine if a scientific article was automatically generated by AI in order to filter out publication spam) or new social protocols (e.g., at what level of contribution might it be appropriate to list an AI as a scientific co-author). 

Many participants took a scientific perspective on defining trust in quantifiable ways. For example, if a science AI system is one that makes predictions about scientific phenomena, then one can measure what percent of these predictions are accurate (and perhaps compare that to a baseline such as predictions by expert humans). This track record will allow scientists to make informed decisions about the extent to which to trust the AI. The notion of earning trust by having a track record of accuracy over a period of time was important to P3, who said, “in the early days I would be gut-checking every single thing [an AI said].” P8 made an analogy to how scientists themselves earn trust over time in the field by building up a strong record of work. P16 also made an analogy to how humans must build up trust over time, noting “I don’t even trust many of my colleagues. Not because I think they have ill intent, but almost anyone I work with will at times have the wrong intuition about something,” implying that scientists should always be skeptical of an AI as well, no matter how many tests of “trust” it can pass.

P3 felt human baselines were important for comparison, such as for an AI that could propose experimental designs - he would want to test if the AI could suggest an improvement over an experiment he had designed himself. Similarly, P10 envisioned experiments in which you set certain scientific tasks and ask the AI and expert scientists to answer this set of questions, and compare their success rate. On the other hand, P6 felt comparisons to physical simulations (rather than humans) might be sufficiently convincing, if AI simulations could produce more accurate predictions than these more traditional simulation systems. P18, on the other hand, indicated a much more formal standard for trust – a formal clinical trial of AI technology, just as other technologies in the medical or biosciences space might be tested. 

P2 took a utilitarian perspective; he felt the notion of “trust” per se was less relevant than the question of whether an AI tool had demonstrable utility (descriptive utility, predictive utility, time-savings utility).  

Some subsets of the scientific community are more likely to be accepting of AI than others. In many fields (including biology, materials science, climate science, and others), there are already divides between experimentalists and those who employ computational approaches. 

Participants felt that including people with domain knowledge in the loop of designing and using Generative AI in the sciences was critical for ensuring scientific accuracy. P8 noted that any scientific AI should include a “human in the loop for an extended period of time… and those people should have domain knowledge.” P9 noted that while he doubted a model could ever replace a physicist, they might “be able to collaborate… on certain tasks.” P11 noted that for medical or clinical applications of AI, human validation was vital for minimizing the serious consequences of false positives or false negatives in diagnostic tasks. He emphasized a need to “get the human values” in the process of scientific discovery and application.

\section{Discussion}
Participants identified many opportunities for Generative AI to impact the sciences, including scientific education, research, and communication. In nearly all cases, the perspective was of AI as a tool that would \textit{complement}, \textit{enhance}, or \textit{accelerate} scientists’ abilities, rather than something that would \textit{automate} the practice of science.

While none of our participants felt that senior experts in their fields such as themselves were likely to be replaceable by AI anytime soon (or ever), P1 (an economist) observed that job displacement in many areas as a result of increasingly powerful Generative AI models was quite likely to cause societal disruption, “AI will hollow out a middle level of workers…”; this may apply to the sciences as well as other areas, with less need for certain types of graduate students, lab assistants, teaching assistants, programmers, etc. Employment disruption in society as a direct result of AI automation or as a result of new scientific discoveries enabled by AI-accelerated science is likely to be highly disruptive in the near- and medium-term, even if ultimately such disruptions result in accelerated scientific progress that benefits society and eventual job reskilling. As P1 observed, “I think the hundred-year answer about technology… is we don’t regret the jobs we lost 100 years ago.”

The tendency of status quo Generative AI tools to make serious factual errors and hallucinations is currently a major obstacle to the use of these tools in any aspect of science. While improved models combined with extensive citation and other explainability features may mitigate this concern, \textit{accuracy} is a key value to this constituency, and will be critical to the success of any future scientific AI. 

Further, it is critical to reflect carefully on the potential negative impacts of future AI tools on the sciences; there is a risk that any gains in productivity or knowledge might be negated by losses resulting from new AI-enabled blights such as cheating or over-automation of scientific writing (and resulting loss of critical thinking skills in the training of future scientists), publication spam (and the resulting impact on scientists’ attention and reviewing load), and fake data and science (and the resulting negative impacts on public trust in science, misdirection of scientific funding, etc.). The potential positives and negatives of particular AI tools on the sciences must be carefully considered, in conjunction with domain experts, as a part of the design and potential deployment of any new AI system for this constituency.

\subsection{Limitations}
While our sample includes twenty scientists with a range of disciplinary affiliations, there are a great many areas of expertise that our study did not touch upon. Additionally, our participants all lived and worked in the United States; participants with differing national/cultural backgrounds might have different perspectives. 

Because all of our participants were affiliated to some extent with Alphabet, they are likely (1) more familiar with Generative AI than other scientists (other than, perhaps, computer scientists), and (2) may be predisposed to more techno-positive attitudes than their peers. For instance, we were surprised that despite not being computer scientists, all of our participants had some degree of familiarity with Generative AI and awareness of deployed systems (e.g., ChatGPT), probably as a result of their exposure to the technology industry through their employment at Alphabet. 

We did not include Computer Scientists among our interviewees since knowledge about the domain of computer science is already quite common among most Alphabet employees; the goal of this work was to gather additional perspectives. 

It is extremely difficult to speculate about future technological capabilities that do not yet exist. As such, participants’ responses regarding potential applications of and concerns about Generative AI are likely heavily influenced by the capabilities of status quo systems and assumptions that future systems are simply faster, more accurate versions of today’s paradigms; this is likely an incorrect assumption. Further, many of participants’ suggestions did not necessarily seem specific to Generative AI models; some capabilities (such as discovering patterns in datasets) may be accomplished with other classes of machine learning, whether status quo or future systems.

Note that some of participants' suggestions and our findings may already be outdated due to the fast pace of progress in Generative AI research. These interviews were conducted in January and February of 2023; the release of the powerful GPT-4 language model in March 2023 \cite{gpt4techreport} already substantially alters the status quo.   

\section{Conclusion}

We interviewed twenty scientists across the physical, life, and social sciences to learn more about how advances in Generative AI might impact scientific professions. Our interviews revealed possible applications of new AI tools across a large swath of scientific practice, including education, data, literature reviews, coding, discovery, and communication. While there is great potential for new AI technologies to augment scientists’ capabilities and accelerate knowledge discovery and dissemination, there is also a need for caution and reflection to mitigate potential negative side-effects of scientific AI tools as well as to prevent intentional misappropriations of new technologies. We are on the cusp of an AI revolution that will touch all aspects of society, including the methods and pacing of scientific discovery.  

\section{Acknowledgements}
We would like to thank our interview participants and also to thank the Google Curie team for suggesting and supporting this inquiry, with particular thanks to Michael Brenner, Subhashini Venugopalan, and Dan Liebling for their comments and feedback. 

\printbibliography

\end{document}